\documentclass[structabstract]{aa}
\usepackage{txfonts}
\usepackage{graphicx}
\usepackage{natbib}

\bibpunct{(}{)}{;}{a}{}{,} 

\begin{document}

\newcommand{\Rsun}{\ensuremath{\textrm{R}_\odot}} 

\title{EUV and HXR Signatures of Electron Acceleration During the Failed Eruption of a Filament}
\titlerunning{EUV and HXR during the Failed Eruption of a Filament}

\author{A. Netzel\inst{1},
   T. Mrozek\inst{1,}\inst{2},
   S. Ko{\l}oma{\'n}ski\inst{1}, \and S. Gburek\inst{2}}
     
\offprints{T. Mrozek, \email{mrozek@astro.uni.wroc.p}}

\institute{Astronomical Institute, University of Wroc{\l}aw, ul. Kopernika 11, 
51-622 Wroc{\l}aw, Poland
  \and Space Research Centre, Polish Academy of Sciences, Solar Physics Division
ul. Kopernika 11, 51-622 Wroc{\l}aw}

\date{Received ..... / Accepted ......}

\abstract {} {We search for EUV brightenings in {{\it TRACE}} 171~{\AA} images and HXR bursts observed during failed eruptions. We expect that if an eruption is confined due to interaction with overlying magnetic structures then we should observe effects connected with reconnection between magnetic structures and acceleration of particles.} {We utilized {{\it TRACE}} observations of three well observed  failed eruptions. Semi-automated method was used to search for abrupt brightness changes in the {{\it TRACE}} field of view. EUV images were compared to HXR spatial distribution reconstructed from {{\it Yohkoh}}/HXT and {{\it RHESSI}} data. The EUV light curves of a selected area were compared to height profiles of eruption, HXR emission and HXR photon spectral index of power-law fit to HXR data.} 
{We have found that EUV brightenings are closely related { to the eruption velocity decrease}, to HXR bursts and to episodes of hardening of HXR spectra. The EUV brightened areas are observed far from the flaring structure, in footpoints of large systems of loops observed 30-60~minutes after the maximum of a flare. These are not `post-flare' loops that are also observed but at significantly lower heights. The high lying systems of loops are observed at heights equal to height, at which eruption was observed to stop. We observed HXR source spatially correlated with EUV brightening only once. For other EUV brightened areas we estimated the expected brightness of HXR sources.} {We find that EUV brightenings are produced due to interaction between the erupting structure with overlying loops. The interaction is strong enough to heat the system of high loops. These loops cool down and are visible in EUV range about 30-60~minutes later. The estimated brightness of HXR sources associated with EUV brightenings shows that they are too weak to be detected with present instruments. However, next generation instruments will have enough dynamic range and sensitivity to enable such observations.}

\maketitle 

\section{Introduction}
The solar corona is filled with plasma frozen in coronal magnetic lines. Structures formed in that way can be both quiet and dynamic. Perfect examples of such structures are coronal loops or filaments (prominences). These features can vary both in size (up to $10^{6}$~km for giant arches) and in temperature ($\ge 10$~MK for post-flare loops). Sometimes they undergo dynamic changes that we believe are due to magnetic reconnection. These changes release great amount of energy leading to the most energetic events in our solar system like filaments eruptions, flares and coronal mass ejections.

\begin{table*}
\caption{Analyzed events}              
\label{tab:events}      
\centering                                      
\begin{tabular}{c c c c c c c c c c}          
\hline\hline                        
\noalign{\smallskip}   
No. & Event & Coordinates & \multicolumn{4}{c}{Associated flare} & \multicolumn{2}{c}{Eruption} & CME\\ \noalign{\smallskip}  
\cline{4-7} 
\noalign{\smallskip}  
& & & GOES class & start & max & end & start & max & \\
\hline                                   
\noalign{\smallskip}
    1 & SOL1999-10-22T09:16 & N20W76 & C4.8 & 09:10 & 09:16 & 09:29 & 09:11 & 09:15 & No \\
    2 & SOL2004-07-14T05:23 & N12W62 & M6.2 & 05:02 & 05:23 & 05:27 & 05:19 & 05:25 & No \\   
    3 & SOL2004-08-13T18:12 & S13W24 & X1.0 & 18:07 & 18:12 & 18:15 & 18:11 & 18:23 & Yes \\  
    \noalign{\smallskip}   
    
\hline                                             
\end{tabular}
\end{table*}

\cite{gilbert2007} proposed classification for filament eruptions based on different amount  of magnetic structure and mass that erupted. Three groups proposed are as follows: 1. \textit{Full} - most ($\ge 90\%$) of filament mass and magnetic structure is erupted. 2. \textit{Partial} - divided into 2 classes A and B. Class A is the eruption of the entire magnetic structure with small amount or even no mass. Class B is the partial eruption of magnetic structure with some or no mass. 3.\textit{Failed} - Neither of mass nor magnetic structure escapes from the Sun. Flares might also be classified in a similar way. Solar flares associated with CMEs are the eruptive flares while flares that occur without CMEs are the confined flares  \citep{sakai1996}. 

There are several possible mechanisms proposed that may lead to failed eruptions. Among them is kink instability \citep{torok2005}, reaching an upper equilibrium \citep{vrsnak2001, green2002}, forces within erupting flux rope \citep{vrsnak1990}, magnetic tension force and momentum exchange with the background plasma \citep{wang2009, archontis2008}, insufficient energy released in the low corona \citep{shen2011} and confinement by the overlying coronal magnetic field \citep{hirose2001, wang2007, liu2008, mrozek2011}

Confining mechanism based on interaction with overlying coronal magnetic field was explored both  theoretically and observationally. The theoretical simulation carried out by \cite{amari1999} showed that  erupting magnetic flux during its evolution could reconnect with the overlying arcade and remain  confined. Numerical experiments showed that evolution of a flux rope (and the whole flux system) after its 
emergence is dependent on ambient magnetic field \citep{archontis2008}.  MHD simulations of a confined filament eruption on 27 May 2002 carried by \cite{torok2005} showed that the decrease of the overlying magnetic field with height is a key factor leading to the failed eruption. Similar results have been reported by \cite{wang2007} who examined 104 X-class flares and found that 10\% of them were confined. The confined events were observed closer to the center of an active region where overlaying coronal magnetic field is the strongest.

Recently \citep{mrozek2011} analyzed one failed eruption that was stopped due to the interaction with the overlying coronal magnetic  field. The interaction was observed as an unusual behavior of the front of the eruption. Namely, the eruption front was observed changing its shape, falling back and side eruptions occurred after. Moreover, radial oscillations of the loops that were pushed to greater heights due to the interaction with the eruption were observed. The EUV brightenings were observed in the foot points of the overlying magnetic structure during the deceleration of the eruption. Such brightenings may be produced by non-thermal electrons accelerated during the episodes of interaction between the eruption and the overlying magnetic field. However, HXR sources spatially correlated with the EUV brightenings were not found. The eruption occurred in the active region with a clear quadruple configuration. Flares and accompanied eruptions in such magnetic configuration were theoretically modeled by \citep{uchida1999, hirose2001}. In this model an eruption may interact with overlying magnetic field and may eventually be stopped.

Assuming that during the evolution of the eruption there are episodes of interaction with overlying field we can expect some signatures of the presence of non-thermal electrons. If non-thermal electrons occur in large, overlying magnetic loops then signatures of their presence have to be observed at some distance from the flare and the associated eruption. In the paper (Paper~I) we present qualitative analysis of the three well observed failed eruptions. The energy balance and quantitative analysis will be presented in the subsequent paper.

\section{The Observations}

\begin{figure}
  \includegraphics[width=9cm]{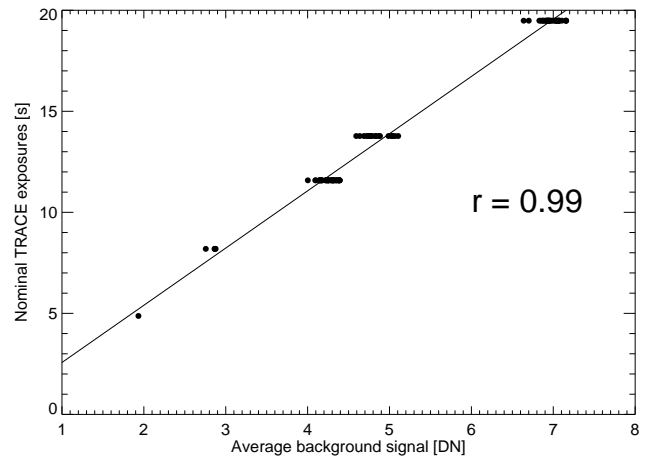}
  \caption{The relationship between TRACE exposure time and average background signal level in images. Average background signal level was determined from the lower left image corner  where the signal was dark, flat and did not change substantially during the observation time interval. Correlation coefficient is shown in the plot.}
  \label{fig:TRACE_exp}
\end{figure}

In our investigation we utilized EUV data from the {\it TRACE} satellite \citep{handy1999} and Hard X-ray (HXR) data obtained by {\it YOHKOH/HXT} \citep{kosugi1991} and {\it RHESSI} \citep{lin2002}. {\it TRACE} was operated since April 2nd, 1998 to June 21st, 2010. The telescope was equipped with a set of filters that enabled us to observe the Sun in several wavelengths covering temperature range from tens of  thousands of Kelvins up to 2 MK. The 30 cm telescope and 1024x1024 CCD detector recorded images with the 1~arcsec resolution (0.5 arc sec pixel size). The typical cadence for images obtained in the same filter was in the range of 2-40 seconds. In our work we used images obtained with 171~{\AA} filter that is sensitive to 1~MK plasma with very small addition of hot (10~MK) component \citep{phillips2005}.

The HXR data from {\it YOHKOH/HXT} and {\it RHESSI} were used to obtain images and spectra. {\it YOHKOH/HXT} was Fourier-synthesis imager equipped with a collimator and a detector plate. The collimator consisted of the grid plates at both ends. The sixty four grids with different slit widths and orientations provide detectors with different Fourier components of actual HXR sources distribution. The detectors measured HXR signals in the four energy channels: L (14-23~keV), M1 (23-33~keV), M2 (33-53~keV) and H (53-93~keV). Images may be reconstructed with MEM or PIXON algorithms. In this work we use PIXON algorithm. The spatial resolution is equal $5\arcsec$. The best time resolution is 0.5~s, but in practice it is worse and depends on count statistics. Since HXR signals were measured in four energy bands it is possible to perform a spectral analysis. The lowest channel, L, may be affected by thermal radiation and should be excluded from the analysis of non-thermal component.

\begin{figure*}[!ht]
\centering
   \includegraphics[width=15cm]{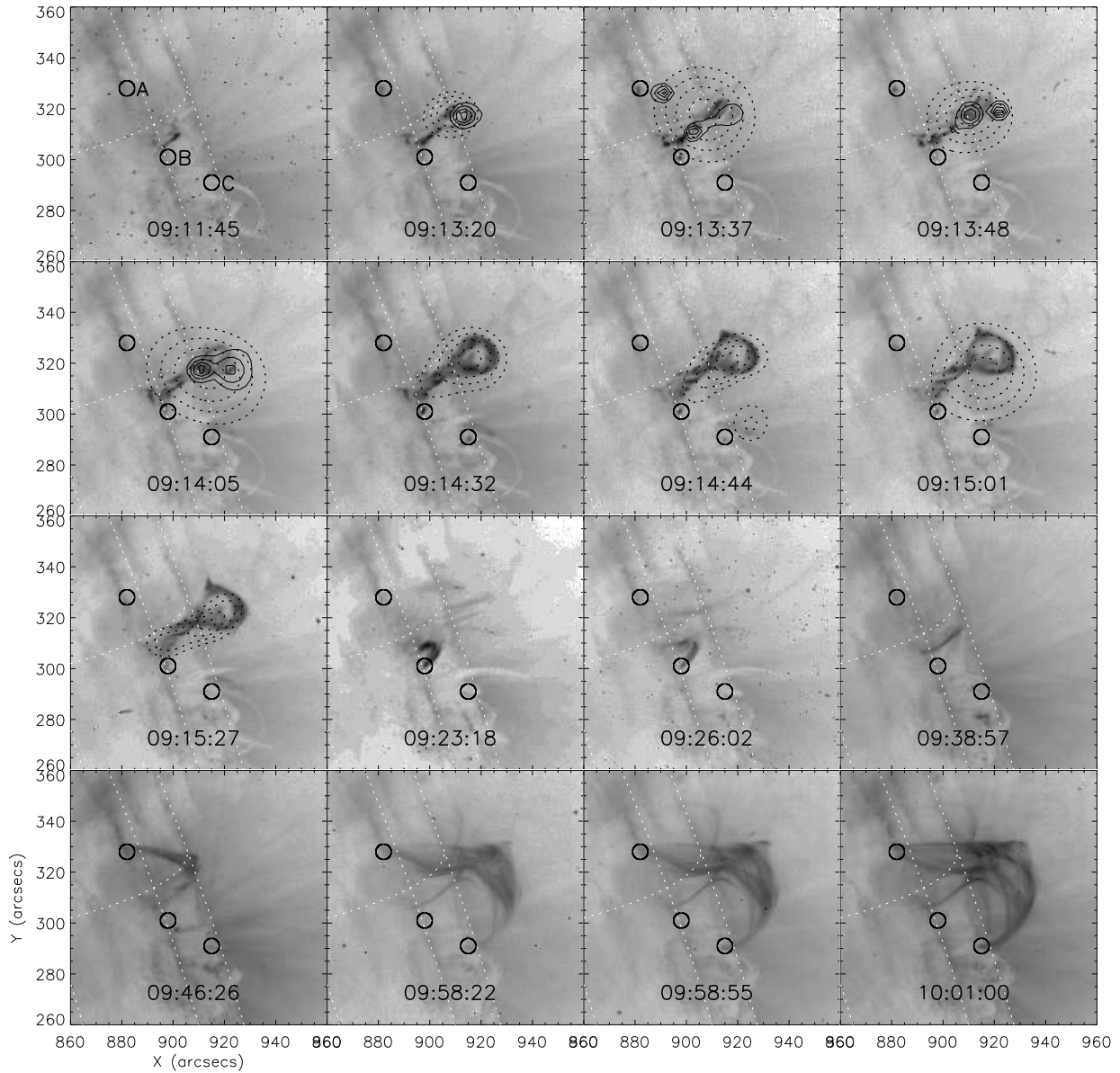}
     \caption{Set of 171~{\AA} images taken from {\it TRACE} (171~{\AA})for different moments during evolution of SOL1999-10-22T09:16 event. Locations of brightenings are circled in every image. In top-left image all regions are denoted (by letters). In selected images reconstructed HXR contours are marked for HXT L (dashed line) and M1 (solid line) for 30, 50, 70 and 90\% of maximum brightness. { See on-line materials for color version of the plot and for a movie presenting the entire event.}}
     \label{fig:19991022_imgs}
\end{figure*}

{\it RHESSI} data were available for the events Nos.~2 and 3 (Table~\ref{tab:events}). {\it RHESSI} is a type of Fourier-synthesis imager that produces modulation of incoming HXR radiation through rotation of the whole instrument. The rotation period of {\it RHESSI} is close to 4~s. Images may be reconstructed with several different algorithms \citep{hurford2002}. Here we used the PIXON algorithm \citep{metcalf1996}. The time resolution for imaging is limited by the rotation period. The spatial resolution depends on the grids chosen for reconstruction and on the reconstruction algorithm. In practice the spatial resolution is rarely higher than $7\arcsec$. {\it RHESSI} uses large germanium detectors to measure time and energy of each detected HXR photon \citep{smith2002}. The energies may be measured from 3~keV (1~keV energy resolution) up to several MeV (5~keV resolution). Thus, the instrument provides us with high energy resolution spectra, which allows performing a detailed spectral analysis. 

The failed eruptions are hard to find with automated methods. The main reasons are the following. (1)~the {\it TRACE} EUV images are full of artificial sources like particle tracks, diffraction structures etc. (2)~the failed eruption changes its shape very fast and may change the  direction of the expansion - the temporal resolution of {\it TRACE} images is not sufficient to analyze these changes. (3)~the braking of the eruption may be smooth or abrupt and may be observed along the whole front or only in its part. For the above reasons we have performed visual inspection of the {\it TRACE} movies database{\footnote{http://trace.lmsal.com/Public/Gallery/Movies/}} and found several examples of failed eruptions. After prompt inspection three of them were classified as eruptions stopped by overlying magnetic field structures. The events investigated are summarized in Table~\ref{tab:events}. 

In the EUV range all the events were observed mainly with the 171~{\AA} filter. The {\it TRACE} data reduction was performed with the standard {\it trace\_prep} procedure available is {\it Solar SoftWare} library. Even after the reduction the {\it TRACE} images pointing is usually incorrect by some arcsec. Thus the pointing should be corrected before any comparison with data obtained with other instruments. We performed the {\it TRACE} pointing correction according to the method described by T.~Metcalf{\footnote{http://www.cora.nwra.com/{$\sim$}metcalf/TRACE/pointing.html}}. We used {\it SOHO/EIT} images obtained with the 171~{\AA} filter as the reference data and calculated the shift of {\it TRACE} images. The corrections were not higher than $14\arcsec$ in both directions. However, the correction is significant and important since the observed brightenings have small areas. 

Searching for brightenings was performed with a semi-automated method. Brightenings were determined basing on a light curve for every pixel on a set of spatially correlated EUV images for the whole event. The main requirement for the pixel to be classified as a brightening was a sudden increase in signal. To reduce number of false brightening pixels (like from cosmic rays) we also demanded that the increased signal should appear for at least two subsequent images and has to be surrounded by other brightened pixels. The group of neighboring pixels constitutes an area of brightening. The cumulative light curves presented further in this paper are calculated basing only on signal from pixels that were marked as an area of brightening.

The similar method for finding brightened pixels was used by \citep{mrozek2004}. The method was effective in searching for impulsive SXR brightenings. However, in {\it TRACE} images we can expect several features  that might be recognized as brightenings but they are not and have to be manually excluded. For this reason in each case we had to decide which of the found brightenings was an actual one. The main problems were connected with solar energetic particles, moving structures and a diffraction pattern. Solar energetic particles cause brightenings that can be classified by automatic methods. For this reason we filtered the maps of brightened pixels and removed structures smaller than 4 squared pixels. Solar energetic particles can also produce lines on images thus we also excluded brightened structures with a width equal 1-2 pixels. Our method of finding brightened pixels has a tendency to find moving structures if they only occur. Such structures were removed manually from the map of brightenings. 

Inspection of the TRACE reduced image cube used for the SOL1999-10-22T09:16 event showed that some corrections for the exposure times are also necessary. Namely, the abrupt jumps of unknown origin in the light curve rates for all pixels were detected. The jumps in rate value were present even in dark featureless portions of the images where only background was observed. We checked the information on image exposure times stored in TRACE FITS file headers and found that these jumps occur when substantial changes in exposure time of the images were made. On the other hand we observed that there is very good linear relationship between average background signal level and TRACE exposure times for many of the images from other analyzed events (see Fig.~\ref{fig:TRACE_exp}). This relationship also allows to determine the exposures directly from background signal. We compared the exposures taken from TRACE  data files with these calculated from the background and found large discrepancies between them for some images. For such images we used calculated exposures. This allowed us to eliminate the jump effects in light curve rates.

The last feature producing artificial brightenings is a diffraction pattern that is always present in {\it TRACE} images \citep{lin2001}. It is especially big problem in the case of very bright areas. The diffraction cross consists of almost exact, but fainter copies of a central source with decreasing brightness. The details of the diffraction cross and {\it TRACE} 171~{\AA} Point Spread Function (PSF) are given in \citep{gburek2006}. If the intensity of a bright source is changing then the diffraction sources also change intensity and are classified as brightened areas. We have examined all the brightened areas and have checked whether they are real or are the result of the diffraction. An example is presented in Figure~\ref{fig:20040714_imgs} (panels from 05:23:09~UT to 05:26:50~UT) showing the situation we observed for SOL2004-07-14T05:23 event. The brightened area B was observed, but it was masked by diffraction sources. For this reason we excluded this event from further analysis.

The next step in our analysis was the estimation of the eruption front measured above some reference point. In each of the three events we observed very well defined area where the eruption was rooted. Thus, we defined the reference point as the base of the eruption. The analyzed expanding structures have well defined leading front thus the errors of height estimations are relatively low. We calculated the uncertainties of the leading front position by measuring the position of the front in several different locations. Next we analyzed the spread of the values and obtained the maximal error to be equal 4~pixels ($2~\arcsec$). { The estimated heights of the leading front of the eruption were fitted with cubic splines. The velocity was obtained through differentiating the fits. The differentiation of noisy data may lead to substantial error magnification. The problem of differentiating noisy data is widely discussed in the literature \cite[]{kontar2005, wen2007, temmer2010}. However, we were interested mainly in estimation of the approximate time intervals of the velocity decreasing. Thus, the problem of error magnification is not as severe because we do not analyze individual minima or maxima in velocity and acceleration.}

The absolute height of the eruption is difficult to obtain since we have only 2D information about 3D distribution of magnetic structures. We chose events located close to the solar limb to minimize the projection effects. Moreover, we can calculate the projection of real structures on the sky plane. Thus, we were able to obtain more realistic heights of erupting structures. The description of the method applied is given in the Appendix.

\begin{figure}
\centering
   \includegraphics[width=8cm,height=18cm]{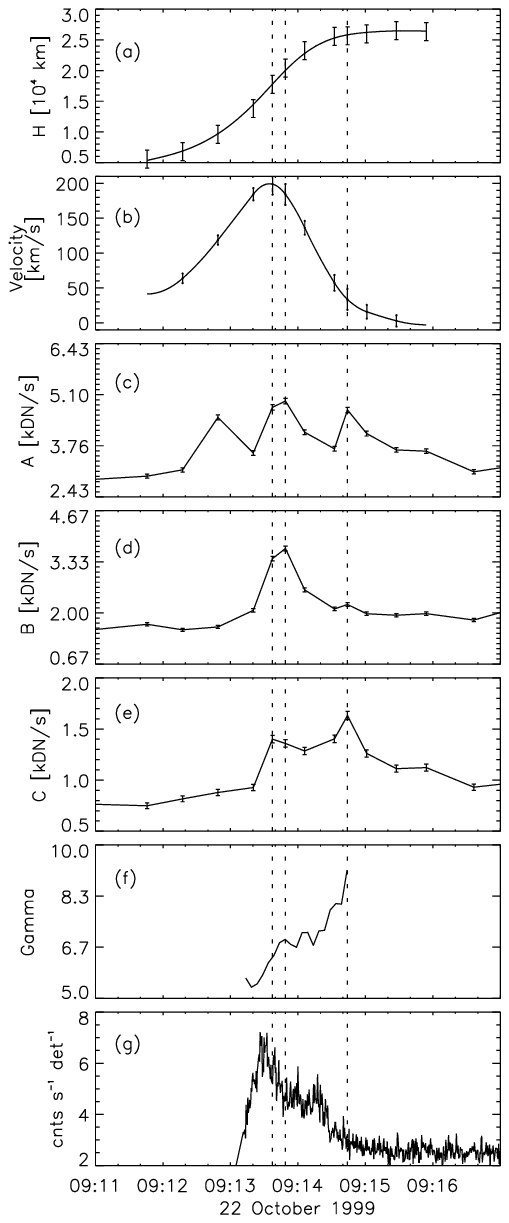}
     \caption{Plots for SOL1999-10-22T09:16 event. { The first two plots show height and velocity against time. The presented measurements of height have been fitted with the use of cubic splines. The velocity is a derivative of the obtained fit.} Next three plots present light curves for brightening regions A, B and C. Next is gamma against time. Last  plot is HXR light curve obtained from HXT instrument aboard {\it Yohkoh} in the M1 channel. In all plots vertical dashed lines represent maxima of light curves for brightening regions. }
     \label{fig:19991022_lc}
\end{figure}

\section{Data analysis}

\subsection{SOL1999-10-22T09:16}
	Figure~\ref{fig:19991022_imgs} shows eruption of a filament at solar limb, which took place on October 19, 1999. The associated flare was classified as C4.8 according to {\it GOES}. For this event there are available observations from \textit{TRACE} at 171~\AA\ and \textit{Yohkoh}/HXT observations for HXR. {\it{EIT}}/LASCO~C2 recorded CME that took place in the same time, which was observed on the east limb. It was not correlated to the analyzed flare since it was visible close to the west limb.
	
The whole evolution of this event is presented in Fig.~\ref{fig:19991022_imgs}. There are no \textit{TRACE} images of the very beginning of the filament rise (between 9:08~UT nad 9:11~UT) available. According to \textit{GOES} the flare begun at 9:10 UT, and the maximal SXR flux was recorded at 9:16~UT. At that moment we also observed the maximal height of the analyzed eruption (Fig.~\ref{fig:19991022_lc}, top panel). During the rising phase the eruption had its greatest velocity around 9:13~UT and at around 9:14~UT it was slowed down. The system of EUV post-flare loops was observed starting from 9:20~UT. They reached the maximum height equal $8\times10^{3}$~km. Moreover, at around 9:40~UT we observed abrupt occurrence of high loops (their height is about $2.8\times10^{4}$~km) that are the separate system, not related to the post-flare loops. They are high and can be compared to the maximal height of the eruption front.
	
We have found three regions that meet the criterion for a chromospheric brightening. They are marked as A, B and C in the top-left panel of Fig.~\ref{fig:19991022_imgs}. EUV light curves for those regions are shown in  Fig.~\ref{fig:19991022_lc} (panels d-f). Their shape resembles the shape of the velocity curve as well as the HXR curve. There are two maxima at 9:13 UT and 9:14~UT observed in EUV light curves. The first maximum correlates well with the maximum in HXR light curve but the second one is about 15 seconds after the corresponding HXR maximum. { The maxima on the light curves are observed simultaneously with slowing down of the eruption. For all the regions the first maximum precedes the moment of the slow down of an eruption.} The two brightening regions A and C are located near the foot points of the high system of loops (Fig.~\ref{fig:19991022_imgs}, panel at 10:01~UT) and region B is located close to the footpoint of the flaring structure (Fig.~\ref{fig:19991022_imgs}, panel at 09:26~UT).

	\begin{figure*}
\centering
   \includegraphics[width=15cm]{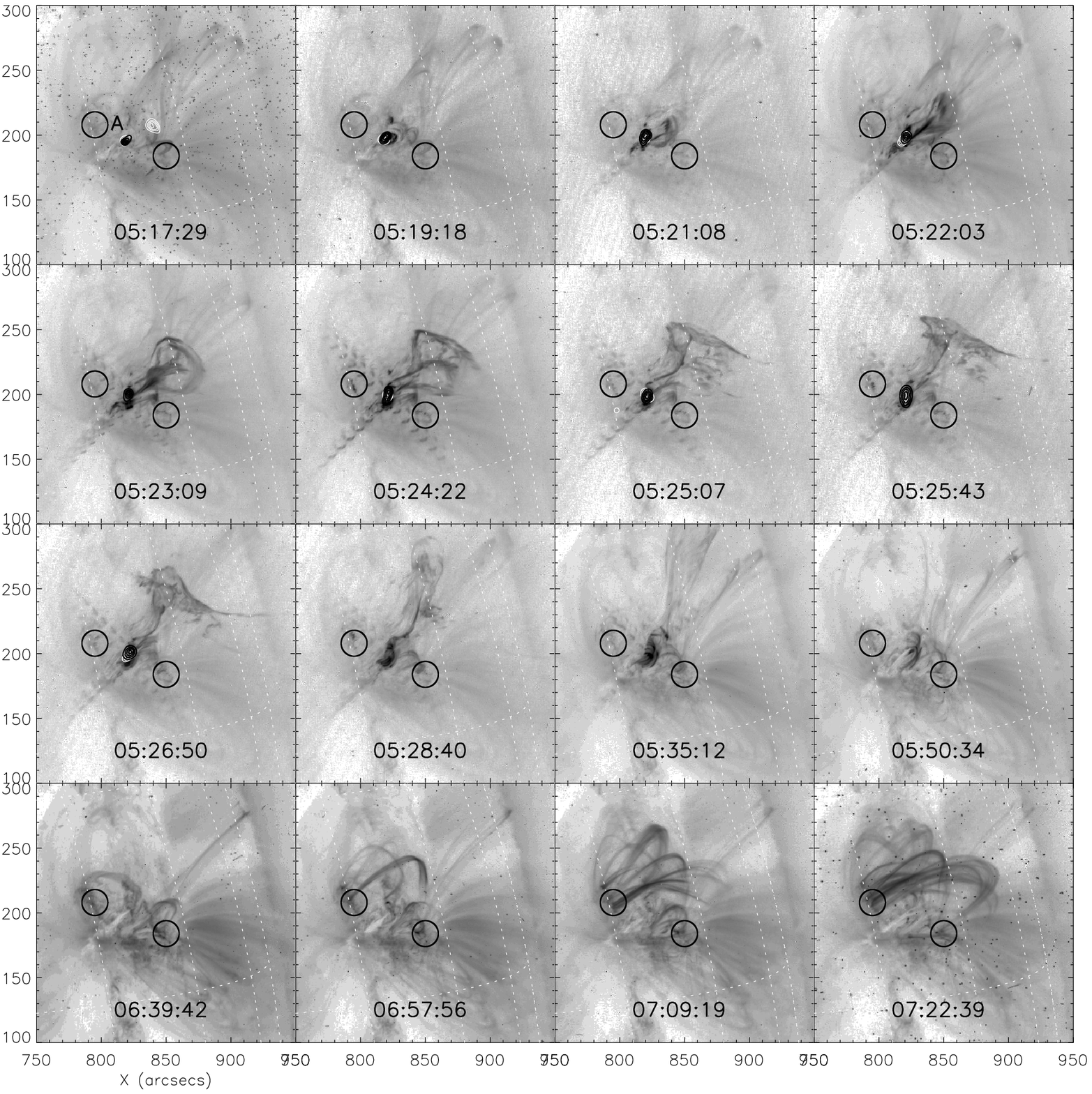}
     \caption{Set of 171~{\AA} images taken from {\it TRACE} for different moments during evolution of SOL2004-07-14T05:23 event. Location of brightening is cricled in every image. In top-left image region is named. In selected images reconstructed HXR contours are marked for {\it RHESSI} energy bands 6-12~keV (white solid line) and 25-50~keV (black solid line). { See on-line materials for color version of the plot and for a movie presenting the entire event.}}
     \label{fig:20040714_imgs}
\end{figure*}

HXR sources were reconstructed for two HXT channels: L and M1. We have chosen time intervals that cover times of {\it TRACE} images. The L-channel source is large and diffuse and is spatially correlated with the evolving eruption. The HXT/M1 channel sources are significantly smaller and their behavior is much more interesting. Namely, the first source is visible several seconds before the maximum burst of HXR radiation (Fig.~\ref{fig:19991022_imgs}) and is located on the top of the eruption front. In the second image we observe two weak, compact sources located lower in the corona. They are spatially correlated with the EUV brightenings A and B. The correlation between the HXT/M1 source and the brightening A (the foot point of the large magnetic structure observed later) may be the result of a non-thermal electron beam. Moreover, there is a weak HXT/L emission source that correlates with the area C (Fig.~ \ref{fig:19991022_imgs}) observed at 9:14:44~UT - the moment of the strongest brightness observed in C. These high loops are invisible for the time of eruption, but can be traced on images taken 30 minutes later. That suggests that we observe the electrons accelerated during the interaction between the eruption front and the system of loops located above. The next image (Fig.~\ref{fig:19991022_imgs}, top-right panel) shows HXT/M1 sources located again within the eruption. One of them is similar to sources observed by \citet{alexander2009} who reported HXR sources in X-point of the eruption. The second source is located at the front of the eruption and is possibly another manifestation of the  reconnection ongoing between the eruption and the overlying high loops. The similar situation (x-point source and at-the-front source) we found for the next moment (09:14:05~UT), but sources are weaker and more diffuse due to the low level of measured signal. { The observed sources occurred during the strongest HXR burst, during the deceleration of the eruption and they are correlated with the occurrence of EUV brightenings.}

\subsection{SOL2004-07-14T05:23}
The second of the analyzed failed eruptions was observed on July 14, 2004 and was accompanied by a M6.2 flare. The analyzed event was observed in an active region NOAA~10646. This event was also analyzed in \cite{mrozek2011} as an example of a failed eruption that started vertical oscillations of overlying coronal loops.

For this event observations from {\it TRACE} (171~{\AA}) and {\it RHESSI} are available. The event SOL2004-07-14T05:23 had no associated CME. According to {\it GOES} the flare started at 5:02~UT and reached the maximum brightness at 5:23~UT. The evolution of the entire event is presented in Figure~{\ref{fig:20040714_imgs}. It can be divided into two phases. The first, between 5:17~UT and 5:24~UT, when the eruption rose up to its maximum height (Fig.~\ref{fig:20040714_lc}). During this phase the highest HXR signal was recorded at the beginning of the eruption rise. The maximum in HXR is correlated with the lowest value of HXR's photon spectral index and { is observed close to the maximum of a velocity}. The end of this phase took place at 5:24~UT when the rising eruption got stopped. At that time the main front of the eruption is held at almost the same height. Next phase started at 5:27~UT when the abrupt change of the main front direction appeared (Fig.~\ref{fig:20040714_imgs}, at 5:28.40~UT). We treated it as a side eruption front and thus it was not included in the analysis of the height. At that moment also small post-flare loops became apparent at the location where the eruption started to rise (Fig.~\ref{fig:20040714_imgs}, panels for 5:35~UT and 5:50~UT). 

From 6:57~UT a system of high loops started to be visible. Again, as for SOL1999-10-22T09:16 event, their height ($5.5\times10^{4}$km) corresponds to the height, at which eruption was confined in the first phase. The high lying system of loops is a structure distinct from the small post-flare arcade (Fig.~\ref{fig:20040714_imgs}, panels from 5:26:50~UT to 5:50:34~UT). The small post-flare loops reach the maximum height equal to $10^{4}$km. Moreover, the foot points of the post flare arcade and the foot points of the high loops are located in  different locations.
			
		There are two regions classified as brightenings (Fig.~\ref{fig:20040714_imgs}, top-left panel), but one of them overlaps with diffraction pattern on {\it TRACE} images. Thus, its emission was affected by the changes of brightness of the foot points of the flare. For this reason this brightening was excluded from the further analysis. The light curve for the remaining region has three distinct local maxima (Figure~\ref{fig:20040714_lc}). { All of them occurred when the systematical decrease of velocity of the eruption was observed.}  They correlate with local drops in HXR photon spectral index. The third maximum took place near the moment when the side eruption started. Region A is located far outside the place where the eruption started to rise. However, it is located at one of the foot points of the high loop system (Fig.~\ref{fig:20040714_imgs}, panel for 7:09~UT). 
		
The entire event was observed by {\it RHESSI} that allowed us to reconstruct images and perform fits to the observed spectra. We reconstructed HXR images in two energy channels 6-8~keV and 14-25~keV. The HXR emission sources are very compact and are spatially correlated with the flare site. We did not record any significant emission located at the front of the erupting structure. However, in a few reconstructed images we observed a very weak HXR source located close to the region A. The emission was below 10~\% of the maximum hence it is not presented in Fig.~\ref{fig:20040714_imgs}. 

The HXR light curves and the HXR photon spectral index evolution show some interesting coincidences with EUV brightenings and kinematics of the eruption. Namely, we do not observe any significant brightenings of the area A during the impulsive phase, which suggests that there is no connection between the flare and the region A. We detected significant changes of brightness of this region during the braking of the eruption front (Fig.~\ref{fig:20040714_lc}). The strong EUV brightenings were accompanied by the very small HXR brightenings (12-25~keV). There are also simultaneous episodes of the spectral hardening. Taking into account the overall behavior we conclude that the area A is a good example of an area that was hit by non-thermal electrons accelerated during the eruption braking due to interaction with overlying field.

\begin{figure}
\centering
   \includegraphics[width=8cm,height=16cm]{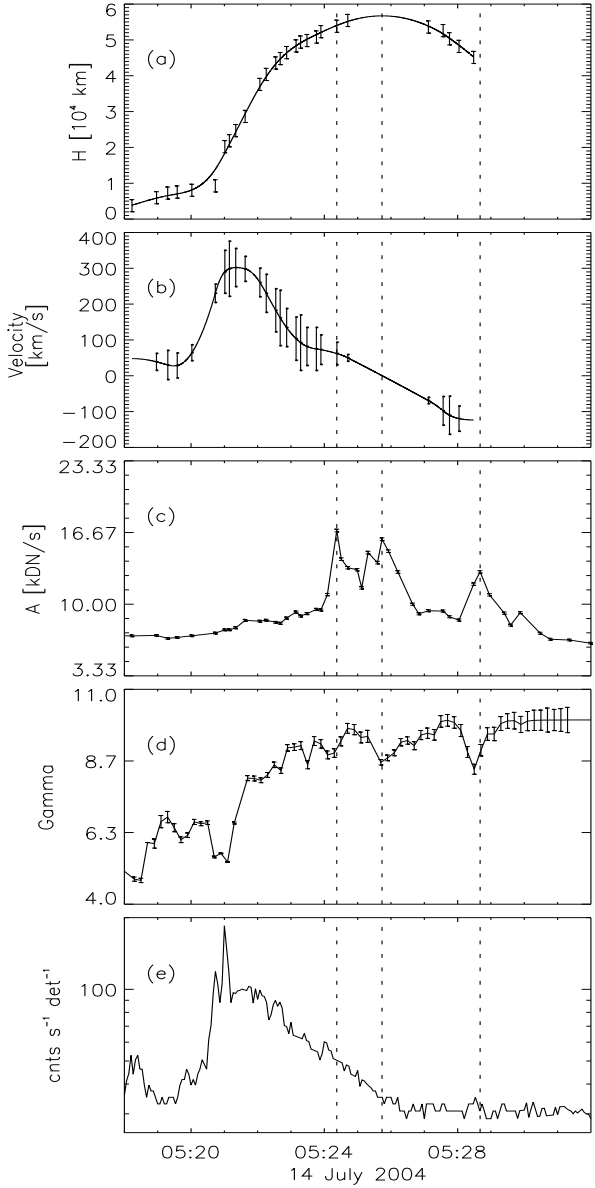}
     \caption{Plots for SOL2004-07-14T05:23 event. { The first two plots show height and velocity against time. The presented measurements of height have been fitted with the use of cubic splines. The velocity is a derivative of the obtained fit.} Next plot presents light curve for brightening region A. Next is gamma against time. Last plot is HXR light curve obtained from {\it RHESSI} energy band 12 - 25 keV. In all plots vertical dashed lines represent three local maxima of light curve for brightening region. }
     \label{fig:20040714_lc}
\end{figure}
 
\subsection{SOL2004-08-13T18:12}

The last of the analyzed events happened on August 13, 2004 in the active region NOAA 10656. According to \textit{GOES} classification it was the X1.0 flare and its maximum flux was observed at 18:12~UT. For EUV range there are \textit{TRACE} (171~\AA) observations. HXR emission was detected by RHESSI for the entire flare. The evolution of this event is illustrated in Fig.~\ref{fig:20040813_imgs}. There was observed a CME associated with this event.
	
\begin{figure*}
\centering
   \includegraphics[width=15cm]{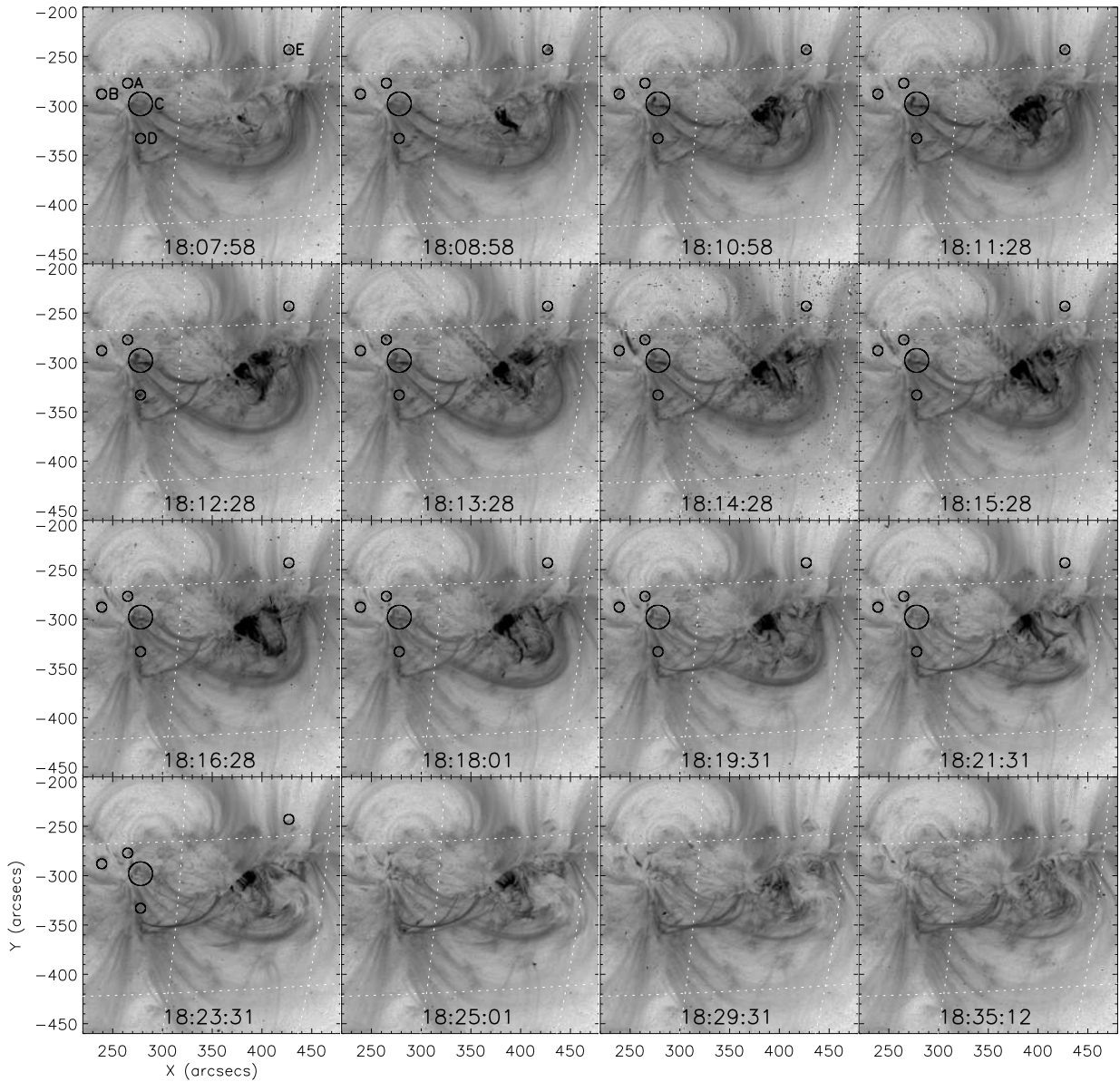}
     \caption{Set of 171~{\AA} images taken from {\it TRACE} for different moments during evolution of SOL2004-08-13T18:12 event. Locations of brightenings are circled in every image. In top-left image regions are named. In selected images reconstructed HXR contours are marked for {\it RHESSI} energy bands 6-12~keV (white solid line) and 25-50~keV (gray solid line). { See on-line materials for color version of the plot and for a movie presenting the entire event.}}
     \label{fig:20040813_imgs}
\end{figure*}

\begin{figure*}
\centering
   \includegraphics[]{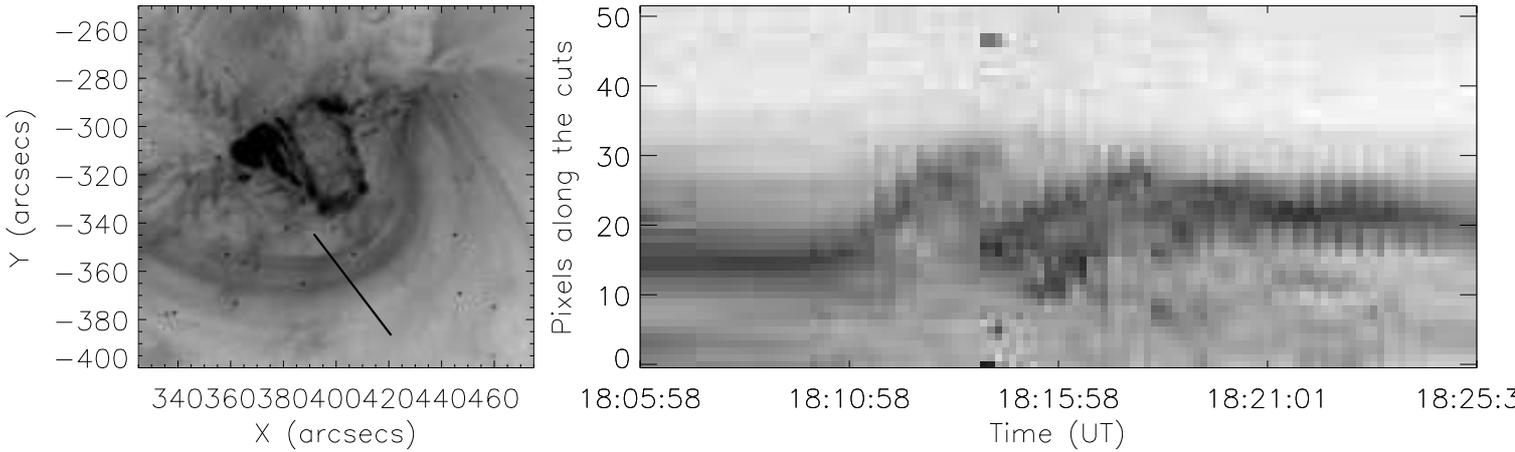}
     \caption{Example of the cut through the overlying system of loops for SOL2004-08-13T18:12 event.}
     \label{fig:20040813_cuts}
\end{figure*}	
	
\begin{figure}
\centering
   \includegraphics[width=8cm,height=18cm]{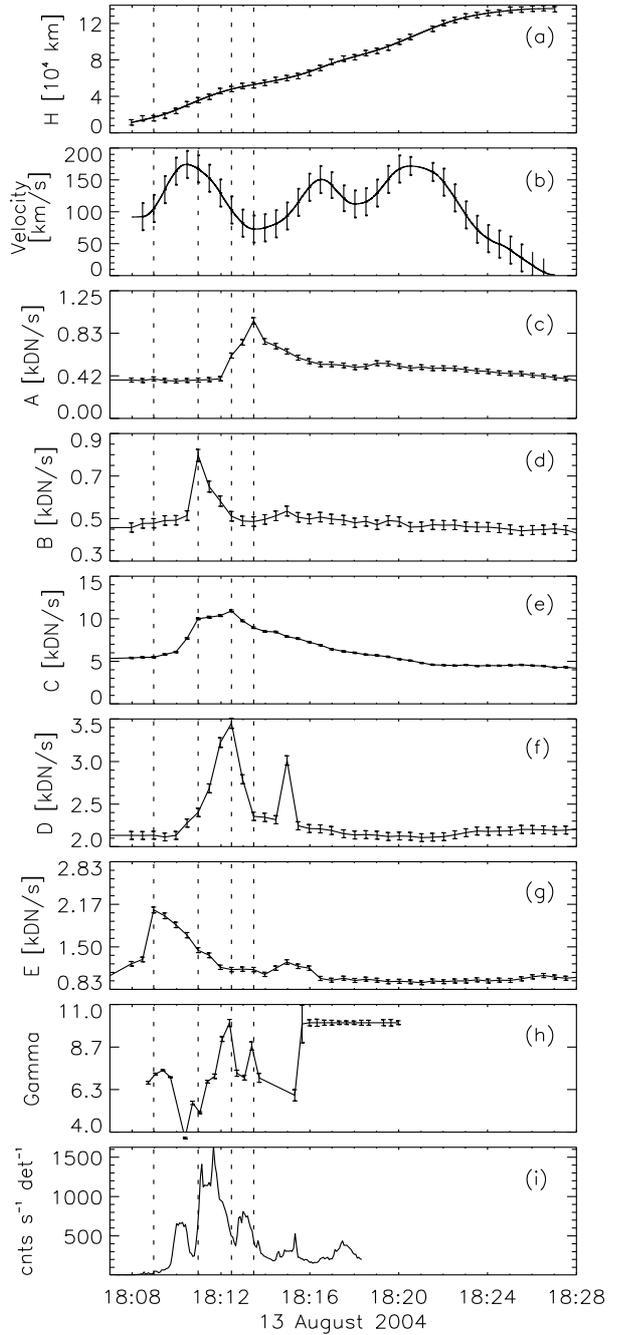}
     \caption{Plots for SOL2004-08-13T18:12 event. { The first two plots show height and velocity against time. The presented measurements of height have been fitted with the use of cubic splines. The velocity is a derivative of the obtained fit.} Next five plots present light curves for brightening regions A - E. Next is gamma against time. Last plot is HXR light curve obtained from {\it RHESSI} energy band 25 - 50 keV. In all plots vertical dashed lines represent maxima of light curves for brightening regions. }
     \label{fig:20040813_lc}
\end{figure}

	The flare begun at 18:07~UT according to \textit{GOES}. In contrast to the previous events the high system of overlying loops is visible even on images taken before the eruption start. The height of these loops is equal to $9.4\times10^{4}$~km before the beginning of the eruption. From 18:10~UT we observed the development of the eruption. Simultaneously the overlying loops were observed stretching  outward as can be seen in Fig.~\ref{fig:20040813_cuts}. The movement of the loops overlying the eruption stopped at 18:21~UT since the eruption  was fully stopped at 18:24~UT. After that typical post-flare loops became visible around the region where eruption started. Their height is equal to $1.2\times10^{3}$~km. 
	
We observed the clear interaction between the eruption and the overlying loops, which is presented in Fig.~\ref{fig:20040813_cuts}. We analyzed the changes of the overlying loops height along a cut that is presented in the left panel of Figure~\ref{fig:20040813_cuts}. The overlying loops were observed to rise together with the evolving eruption. Obviously, we have only 2D image and there is no exact information about the direction of this movement. It is possible that we observe an eruption that elbows overlying structures and there is no stretching. If the overlying loops were pushed to a side then, due to projection, we  should observe initial lowering of the loop tops. We did not observe such a behavior (Fig.~\ref{fig:20040813_cuts}, right panel) and treated it as a suggestion for rising height. 
	
We found five regions that met the chromospheric brightening criteria. They are labeled from A to E  (Fig.~\ref{fig:20040813_imgs}, top-left panel). Their light curves are presented in Fig.~\ref{fig:20040813_lc}. Regions A to D are located near one of the foot points of the overlying high loops. Only region E is located close to the second foot point but it is probably a foot point of other system of loops. The reconstructed HXR contours (6-8~keV and 25-30~keV) are spatially correlated with the flaring structure. We tried to reconstruct HXR sources with different time intervals and energy ranges but we did not find any significant emission correlating with neither the found EUV brightenings nor the eruption. 

Each of the regions brightened at different moment in time but all of them slightly precede the moment when the overlying magnetic loop ceased to move. The light curve for the region C shows an increase in signal at the moment of the maximum in the HXR curve (Fig.~\ref{fig:20040813_lc}). The intensity of this region initially rises and stays almost at the same level between 18:11~UT and 18:13~UT. After this time the intensity is slowly droping to the background level. The light curve of the region A has a similar shape, but its maximum is closer to the second highest HXR peak at around 18:13~UT. The light curve of the region D has two distinctive maxima and the first one is well correlated with the minimum of the HXR photon spectral index. The light curve of the region B has its maximum near the highest HXR maximum. The light curve of the region E shows weak reaction correlated with the HXR peak around 18:15:30~UT. All of them happened in the beginning phase of the eruption. The last brightening was observed around 18:13~UT. { After that time we observed next phase of velocity rising.} Such a behavior strongly supports the scenario, in which the brightenings occur during episodes of interaction between the eruption and the overlying large system of loops. 

\section{Results}

\subsection{Time correlations}

We tried to search for correlations between the kinematics of the eruption, EUV brightenings, HXR emission and HXR photon spectral index. In all three cases we have found that such a relation exists. Usually, we observed the HXR burst first and shortly after that we detected the local minimum of HXR photon spectral index. { Simultaneously we observed a decrease in velocity.} These features were usually observed just before or simultaneously with EUV brightenings. The EUV brightenings were never observed before the HXR burst and they were correlated to the decrease of the eruption velocity. The time shift observed was usually less than several seconds. 

In the case of SOL2004-07-14T05:23 we observed a series of EUV brightenings that occurred during the long (about 6~minutes) stage of the eruption braking. These brightenings were observed when HXR emission (12-25~keV) was on the decay phase. However, we measured a significant hardening of the HXR spectra, which is well correlated in time with EUV brightenings. Thus, we expect that these brightenings are caused by non-thermal electron beams. 

The SOL2004-08-13T18:12 event was the one, in which we observed overlying loops since the early phase of the eruption evolution. The EUV brightenings were observed only during the initial phase of the eruption. It was the  phase when the overlying loops were observed changing their height also (Fig.~\ref{fig:20040813_cuts}). At almost the same time when the overlying loops reached their maximum height we observed the last EUV brightening in the foot point of these loops. After that we still measured an increase of the eruption. It is possible that the eruption pushed through the overlying field and the interaction with overlying field finished. The observational signature for such a scenario is the associated CME - clearly visible but very slow and only a few degrees wide.

\subsection{Spatial Distribution of Found Brightenings}

In all the events presented we have found EUV brightenings that were observed during the evolution of the failed eruptions. The majority of these brightenings was spatially correlated with large magnetic structures that  enclose the eruption. These large systems of loops (height up to $6\times10^4$~km) are not typical post-flare loops. We observed system of post-flare loops that evolved with continuously rising height but, in the analyzed events their maximum height was less than $10^4$~km. Analyzing occurrence of post-flare loops we have found that there is discontinuity in heights between low-lying and high-lying loops (see the movies in on-line material). It suggests that there is a weak relationship between them. Moreover, the EUV brightenings observed are located in foot points of these large loops. The analyzed EUV brightenings are not flare foot points and are observed far from the flare site. A typical distance between the flare site and the EUV brightenings is larger than $2\times10^4$~km. 

The large system of overlying loops is in close relationship with maximal height reached by the eruption. The observations suggest that there is an interaction between the eruption and the overlying structures. For example in SOL2004-07-14T05:23 event we observed the change of front's shape at heights similar to heights of the large loops visible an hour after the flare. In the case of SOL1999-10-22T09:16 we observed the 23-33~keV emission sources atop the eruption, which suggests some episodes of energy release. In the SOL2004-08-13T18:12 we observed the overlying system of loops that were visible during the overall evolution of the eruption and the overlying loops changed height according to the eruption.

We have found the correlation between the HXR emission source and the EUV brightening in one case only (SOL1999-10-22T09:16). However, some of reconstructed HXR images suggest that such sources may exist but they are very weak and are strongly masked by flare emission.

\begin{figure}
  \resizebox{\hsize}{!}{\includegraphics{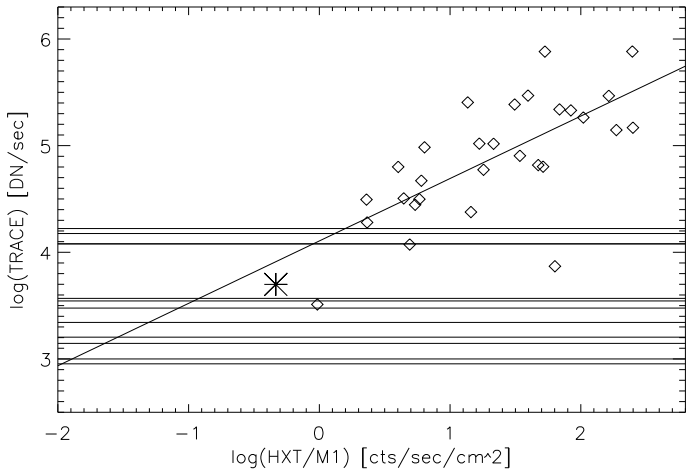}}
  \caption{Correlation between {\it TRACE} 171~{\AA} and HXT/M1 signals. The results obtained by \citet{mrozek2007} are presented with diamonds. The linear fit to these points is presented also. One point estimated for the source A from the SOL1999-10-22T09:16 event is presented with asterisk. Horizontal lines present brightness levels obtained for other EUV brightenings analyzed in this paper.}
  \label{fig:t171_hxr}
\end{figure}

\subsection{Estimation of HXR Sources Brightness}

Temporal characteristics of the brightenings observed show that the time shift between a HXR peak and the maximum of EUV brightenings is very short. It suggests that we may exclude mechanisms like waves or mass draining  operating in different time scales. Thus, we have to answer the question: why do not we see spatially correlated HXR sources if the EUV brightenings observed are of the non-thermal electrons beam origin?

To answer this question we moved to the analysis made by \citet{mrozek2007} where EUV and UV brightenings observed in flare foot points were compared with HXR emission sources. The authors selected a group of flares with well developed impulsive phases and clearly observed foot point sources. The analysis was based on {\it TRACE} and {\it Yohkoh}/HXT data. {\it TRACE} foot points are usually saturated during the impulsive phase. The authors developed a method for saturated signal recovery based on the {\it TRACE} diffraction pattern characteristics \citep{gburek2006}. The EUV signal recovery allowed the quantitative analysis of EUV and HXR signals. The results obtained showed that the signals were correlated and we might conclude that both types of sources were caused by non-thermal electrons. The correlations between the EUV (171~{\AA}) and the HXR (HXT/M1  channel) fluxes are presented in  Fig.~\ref{fig:t171_hxr}.

\citet{tomczak1997} analyzed SXR brightenings in flare foot points and compared them to HXR emission sources. He concluded that the observed correlations may be used for analysis of non-thermal electron stream if no HXR sources are observed. According to this we decided to use the correlations obtained by \citet{mrozek2007} to estimate the expected intensities of HXR sources spatially correlated with the observed EUV brightenings. For the events analyzed we have found that none of brightenings was saturated thus we do not need to perform any signal recovery. The EUV intensities of brightenings analyzed in this paper are marked as horizontal lines (Fig.~\ref{fig:t171_hxr}). One point representing the brightening from SOL1999-10-22T09:16 event is presented with an asterisk. All horizontal lines cross with dependence obtained for flare foot points in the left side of a plot. It means that hypothetical HXR sources that may be spatially correlated with the analyzed brightenings are very weak. They are at least 10 times fainter than faintest foot point sources observed with the HXT. However, the main problem is not the intensity of the sources, but rather the dynamic range. In Fig.~\ref{fig:t171_hxr} there is a group of horizontal lines  located at the level that gives expectation for HXR source detection. Unfortunately, at that time HXR images were dominated by bright sources connected with a flaring structure. 

{

\subsection{Heating and cooling the high-lying system of loops}

Foot points of large system of loops observed in EUV are dim and in HXR are even undetectable. This means that the number of electrons accelerated during the interaction between an eruption and higher loops is low. Thus the following question may occur: Is the energy deposited in foot points large enough to evaporate chromospheric material, i.e. heat the material up to 1~MK and transport it high into the corona?

We can answer the question by doing simple estimation for the event No. 1. The large system of loops is observed in the \textit{TRACE}~171~{\AA} band, i.e. the plasma should have temperature about 1~MK. The height and volume of the loops can be estimated also from the \textit{TRACE} images. Densities of a typical EUV loop is about $10^{9}$cm$^{-3}$ \citep{aschwanden1999}. Based  on these parameters we calculated that the total thermal and gravitational energy of the plasma contained in the loops is about $5\times10^{26}$~erg.

On the other hand energy deposited by non-thermal electrons in the HXR foot point of the event No.~1 can be calculated based on a reconstructed HXT/M1 source and correlations between HXR and EUV signals and power-law index of the HXR spectrum obtained by \cite{mrozek2007}. We estimated that the energy is about $10^{28}$~erg. This is more than the total thermal and gravitational energy contained in the large EUV loops. The energy deposited by non-thermal electrons is high enough to heat chromospheric material up to $6-7$~MK and to rise it up to fill the whole large system of loops with plasma of density of about $5\times10^{9}$cm$^{-3}$. The loops filled with such plasma should be visible in SXR. 

The observations for the SOL1999-10-22T09:16 event confirm this expectation. We carefully inspected {\it Yohkoh}/SXT data and found a few SXT/Al.1 (maximum of thermal sensitivity 5~MK) images that partly overlap the analyzed {\it TRACE} images. Namely, the southern leg of the high lying loop is visible close to the edge of the SXT/Al.1 frame. The temperature response of the SXT/Al.1 filter has the maximum close to 5~MK, which is consistent with the estimation made above. The part of the leg is visible between 9:25~UT and 9:43~UT in SXT/Al.1 images. In the full-Sun image taken with the AlMg (maximum of thermal sensitivity 6~MK) filter on 9:04:55~UT there is no such structure. 

If we assume that heating the loops to 5-7~MK due to an interaction may occur then other problem may arise. Namely, the time difference between the eruption stop and the occurrence of high lying loops is of the order of 40-90 minutes. If we interpret these loops as a system that was heated by the interaction with the eruption and is cooled down after, then the loops should be seen much earlier, assuming typical values of cooling times \citep{antiochos1978}. 

However, recent observations \citep{winebarger2003} show that we must be careful when analyzing the visibility of EUV loops and relating them to hydrostatic cooling times. \cite{winebarger2003} analyzed five loops observed by {\it TRACE} with EUV filters. The authors estimated the time of the appearance of a loop in a given filter. The estimated difference of the appearance in {\it TRACE} 195~{\AA} and 171~{\AA} filters varies from 150~s to 11000~s. In one case (loop~2 in \cite{winebarger2003}) observations made in three EUV filters were available. The time difference of loop visibility for the 284~{\AA} and 171~{\AA} filters was about 50 minutes. The maxima of the temperature response of these filters are 2~MK and 1.3~MK respectively. It means that the loops in the 171~{\AA} filter may be visible even an hour after the heating to a few MK temperatures. 

With the given set of data we are not able to resolve the temperature evolution of the high lying loops with time. We have found a few Yohkoh/SXT/Al.1 images for the SOL1999-10-22T09:16 event. The images overlap the southern part of the analyzed high loops. Thus, not the whole structure can be observed. Nevertheless, the southern leg is visible in the SXT/Al.1 images between 9:25~UT and 9:43~UT. Moreover we inspected SOHO/EIT 195~{\AA} images. The high lying system of loops is visible starting from 9:48~UT but we must keep in mind that the temporal resolution of EIT data is 12~minutes.

The interaction between an eruption and overlying magnetic structures is a complex process. We have to remember that estimated energy contained in non-thermal electron is only a part of the energy released during the interaction. Thus the detailed numerical simulations are needed to explain different features of the interaction.}

\section{Discussion and Conclusion}

\begin{figure}
  \resizebox{\hsize}{!}{\includegraphics{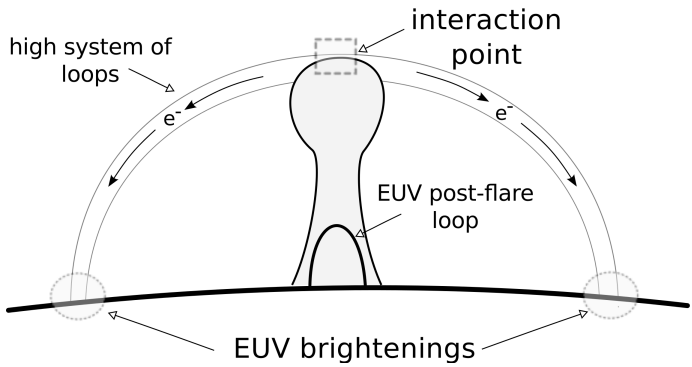}}
  \caption{The scenario suggested for the production of EUV brightenings analyzed in this work.}
  \label{fig:scenario}
\end{figure}

In this paper we present observations of three failed eruptions recorded by {\it TRACE} with a use of the 171{\AA} filter. EUV observations are supported by HXR data from {\it Yohkoh}/HXT and {\it RHESSI}. We used a semi-automated method to search for regions that abruptly brightened in the {\it TRACE} images. Regions that were classified as EUV brightenings are usually located far from flare loops so they are not directly correlated with processes going on there. 

We have found some common behavior of brightenings. Both, temporal and spatial characteristics, may be summarized as follows:

\begin{itemize}
	\item { The characteristics analyzed (height, velocity, HXR photon spectral index, EUV brightenings, HXR ligh curves) are time correlated. Usually an HXR peak and the minimum of a HXR photon spectral index are observed a few seconds before an EUV brightening.} 
	\item { The maximum EUV intensity of selected regions was observed during the deceleration of an eruption front.} 
	\item The EUV brightenings were found with a use of a semi-automated method. They are in exact positions as the observed footpoints of the large systems of loops that are observed tens of minutes after the eruption braking episodes. These loops are not typical 'post-flare loops'. 
	\item The heights of the large system of loops are almost the same as the maximum height reached by the eruption. Together with the fact that the brightenings are observed in footpoints of these loops, it suggests that the loops exist since the beginning of the eruption, but for some reasons (temperature, density) are not visible. 
	\item The interaction with the eruption causes the heating of the loops and, after being cooled down, 30-90~minutes later the loops become visible in EUV. { \cite{warren2002} explained such long cooling times assuming that loops have unresolved, filamentary structure. Multi-stranded loops can persist much longer than predicted with a characteristic cooling time. More hypothesis are possible \citep{winebarger2003} like residual heating or non-ionization equilibrium, but these require detailed hydrodynamical simulations.}
	\item In one case we detected very weak HXR emission (23-33~keV) located exactly at the front of the eruption. It may suggest that some energy is released at the places of interaction between the eruption front and the overlying magnetic field structures.
	\item Expected HXR sources connected with the EUV brightenings are weak and may rarely be observed with the present instruments. They are about 10 times fainter than faintest foot points observed.
\end{itemize}

Basing on the available data, we present the following common scenario for the observed events (Fig.~\ref{fig:scenario}). It have to be noted that above scenario should not be confused with so called unified model \citep{shibata1999} of a solar flare. The rising eruption, after the initial acceleration, is slowed down by the interaction with overlying magnetic field. Except of SOL2004-08-13T18:12 the overlying field is not directly visible. In the points where the interaction between the erupting structure and the overlying magnetic loop happens, particles are being accelerated to non-thermal velocities. Next, the particles propagate through probably almost empty magnetic loops (events Nos.~1 and 2) down to the transition region and the chromosphere where they collide with dense matter and produce EUV brightenings. If the chromospheric evaporation occurs then the loops are filled with plasma and after cooling to 1~MK they became visible in {\it TRACE} 171~{\AA} images. 

The importance of overlying magnetic field in the confining eruptions was discussed in numbers of theoretical and observational papers. The results presented here show that the interactions between an eruption and overlying field may be traced in EUV and HXR ranges. In the forthcoming paper we will present a quantitative analysis of these observations and we will investigate energy balance of the observed features.

\section*{Acknowledgements} 
 The {\it RHESSI} and {\it TRACE} satellites are NASA Small Explorer (SMEX) missions. The {\sl Yohkoh} satellite is a project of the Institute of Space and Astronautical Science of Japan. We acknowledge financial support from the Polish National Science Centre grants 2011/03/B/ST9/00104 and 2011/01/M/ST9/06096  and  from the European Commission's Seventh Framework Programme under the grant agreement No. 284461 (eHEROES project). We acknowledge many useful and inspiring discussions with  Professor Micha\l \mbox{ }Tomczak. We also thank Barbara Cader-Sroka for editorial remarks. We are grateful to the anonymous referee for fruitful discussion and many remarks that helped to improve the paper.

\appendix
\section{Estimation of absolute height of eruptions}

\begin{figure*}
\sidecaption
\includegraphics[width=12cm]{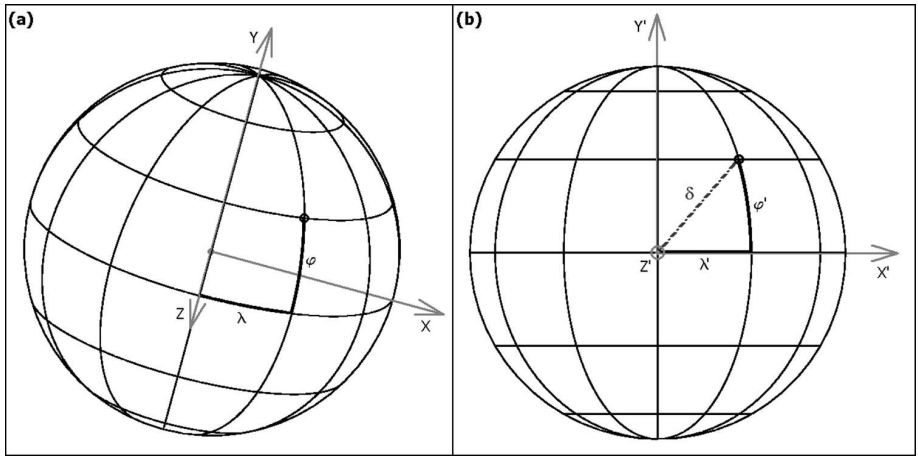}
\caption{Heliographical and Cartesian coordinate systems used to corrected altitude/height for the projection effect: (a) real 
heliographical coordinate system $(\Lambda \Phi)$ and related Cartesian system ($XYZ$), (b) rotated heliographical coordinate 
system $(\Lambda \Phi)'$ and related Cartesian system $(XYZ)'$. The $z'$-axis directed to an observer.}
\label{f1}
\end{figure*}

The height of an eruption above the photosphere is a very important parameter. It helps to understand kinematic evolution of this phenomenon. 
All the heights estimated in this paper were corrected for the projection effect using a method described below. The method allows to estimate real altitude/height of any source/feature observed in the solar corona.

To estimate the real height of an eruption we need to know its position in the \textit{TRACE} image and find the point on the solar photosphere, above which the eruption is situated (reference point). The eruption's position was defined by the position of its leading front. The position of the reference point was taken from locations of footpoint sources visible simultaneously with the eruption in the \textit{TRACE} images.
 
For a single eruption we have several \textit{TRACE} images, in which the eruption was visible. The position of the leading front of the eruption was measured in each available image. The position of the reference point was taken from only one image, in which foot point sources were the most pronounced. Since the time, for which the position of the reference point was determined is somewhat different from the time of the leading front measurements, all the positions were corrected to account for the solar rotation. For each pair of positions measured, leading front -- reference point, we are now able to estimate the corrected height of an eruption. 

In the first step an apparent distance, $d_a$, between the leading front position and the centre of the Sun was calculated. To correct the apparent distance for the projection effect we need to know an angle, $\delta$, between the direction to an observer and the local vertical direction passing through the reference point. If the solar rotation axis was perpendicular to the observer/Earth's orbit the Sun and the heliographical coordinate system would be visible to the observer as shown in Fig.~\ref{f1}b. In such a case $\delta$ could be calculated directly from the heliografical coordinates of the reference point (longitude, $\lambda'$ and latitude, $\phi'$).

\begin{equation}
\label{e1}
\cos \delta = \cos \lambda' \cos \phi' ~,
\end{equation}

Unfortunately, the rotation axis of the Sun is inclined to the observer/Earth's orbit at 82.75 degrees. Due to this fact at different points of the Earth's orbit, the Sun and heliographical coordinate system are visible at different inclinations, e.g. as shown in Fig.~\ref{f1}a. At any time the inclination of the solar rotation axis as seen on the sky plane can be described by two angles: $\alpha$ -- an inclination towards eastern/western solar limb and $\beta$ -- an inclination towards/outwards the observer. For example in Fig.~\ref{f1}a the Sun is inclined with its north pole toward the observer and toward western limb. An accurate correction for the projection effect demands including $\alpha$ and $\beta$ in calculation of the $\delta$ angle. The easiest way to do this is to convert real coordinates of the reference point ($\lambda$, $\phi$) to a new idealized heliographical coordinate system, such as this shown in Fig.~\ref{f1}b. For this purpose let us define two heliographical coordinate systems and two related Cartesian coordinate systems, centers of which are at the center of the Sun. 

\begin{itemize}

\item The real heliographical coordinate system $(\Lambda \Phi)$ and a related system $(XYZ)$. The $z$-axis of the system is crossing solar photosphere in the point $(\lambda=0, \phi=0)$, the $x$-axis -- in the point (0, 90W) and the $y$-axis agrees with the solar rotation axis (see Fig.~\ref{f1}a).

\item A rotated (idealized) heliographical coordinate system $(\Lambda \Phi)'$, in which the point $(\lambda'=0, \phi'=0)$ is situated in the center of the solar disk. A related system $(XYZ)'$ has the $z'$-axis directed towards the Earth/observer. The $x'$-axis and the $y'$-axis lay in the sky/image plane. Thus, these axes constitute a heliocentric coordinate system that is used to measure positions of any point in e.g. a \textit{TRACE} image (see Fig.~\ref{f1}b).

\end{itemize}

In order to change over from the system $(\Lambda \Phi)$ to the system $(\Lambda \Phi)'$, we have to rotate the first one by the angle $\beta$ about the $x$-axis to agree $z$-axis with $z'$-axis, then rotate by the angle $\alpha$ about the $z$-axis to agree $x$-axis and $y$-axis with $x'$-axis and $y'$-axis respectively. Equations that allow to convert Cartesian coordinates of a point from the system $(\Lambda \Phi)$ to the system $(\Lambda \Phi)'$ are as follows:

\begin{eqnarray}
\label{e2}
x' = (z \sin \beta + y \cos \beta) \sin \alpha + x \cos \alpha ~,   \nonumber \\
y' = (z \sin \beta + y \cos \beta) \cos \alpha + x \sin \alpha ~, \\
z' = z \cos \beta - y \sin \beta ~,   \nonumber
\end{eqnarray}

where $(x, y, z)$ are coordinates of a point in the system $(XYZ)$ and $(x', y', z')$ -- in the system $(XYZ)'$. Since coordinates of points 
on the 'surface' of the Sun are given in the heliographical coordinate system, we must convert Eq.~(\ref{e2}) to this system. Using 
equations:

\parbox{4cm}{\begin{eqnarray*}
x = R_\odot \cos \phi \sin \lambda ~, \\
y = R_\odot \sin \phi ~, \\
z = R_\odot \cos \phi \cos \lambda ~,
\end{eqnarray*}}
\hfill
\parbox{4cm}{\begin{eqnarray}
\label{e3}
x' = R_\odot \cos \phi' \sin \lambda' ~,  \nonumber \\
y' = R_\odot \sin \phi' ~,  \\
z' = R_\odot \cos \phi' \cos \lambda' ~,  \nonumber
\end{eqnarray}}

where $R_\odot$ is the Sun's radius, $(\lambda, \phi)$ and $(\lambda', \phi')$ are co-ordinates in the systems $(\Lambda, \Phi)$ and $(\Lambda, \Phi)'$, respectively, after simple transformations we obtain:
\begin{eqnarray}
\label{e4}
\tan \lambda' = \frac{a_{1} \sin \alpha + a_{2} \sin \alpha + a_{3} \cos \alpha}
                     {\cos \phi \cos \lambda \cos \beta - \sin \phi \sin \beta} ~,  \\
\sin \phi' = a_{1} \cos \alpha + a_{2} \cos \alpha + a_{3} \sin \alpha ~,           \nonumber
\end{eqnarray}

where
\begin{eqnarray}
\label{e5}
a_{1} = \cos \phi \cos \lambda \sin \beta ~,   \nonumber \\
a_{2} = \sin \phi \cos \beta ~, \\
a_{3} = \cos \phi \sin \lambda ~.   \nonumber
\end{eqnarray}

The angles $\alpha$ and $\beta$ for any date/time can be estimated using the SSWIDL procedures. 

Using Eq.~\ref{e4} we can convert measured ($\lambda, \phi$) coordinates of the reference point into much suitable ($\lambda', \phi'$) coordinates and then calculate the angle $\delta$ from Eq.~\ref{e1}. Now, the apparent distance, $d_a$, between a leading front position of an eruption and the centre of the Sun can be corrected for the projection effect. A real distance, $d_r$, can be calculated from the following equation (after including Eq.~\ref{e1}):

\begin{equation}
\label{e6}
d_r = \frac {d_a} {\sin \delta} = \frac {d_a} {(1 - \cos^{2} \lambda' \cos^{2} \phi')^{2}}~,
\end{equation}

Finally, subtracting $R_\odot$ from a distance $d_r$, a real height, h, of an eruption above the photosphere can be obtained: $h = d_r - R_\odot$.

The real height uncertainty is caused mainly by uncertainties in measurement of positions of the eruption leading front and the reference point. We calculated the error on the height, $h$, using standard propagation of uncertainty methods.

\bibliographystyle{aa} 
\bibliography{netzel_mrozek} 

\end{document}